# Toward Compact Data from Big Data[1]

Song-Kyoo (Amang) Kim
Computing Program, School of Applied Sciences,
Macao Polytechnic Institute,
R. de Luis Gonzaga Gomes, Macao, SAR
amang@ipm.edu.mo

*Abstract*—**Bigdata is a dataset of which size is beyond the ability of handling a valuable raw material that can be refined and distilled into valuable specific insights. Compact data is a method that optimizes the big dataset that gives best assets without handling complex bigdata. The compact dataset contains the maximum knowledge patterns at fine grained level for effective and personalized utilization of bigdata systems without bigdata. The compact data method is a tailor-made design which depends on problem situations. Various compact data techniques have been demonstrated into various data-driven research area in the paper.**

*Keywords-Bigdata; compact data; data reduction; machine learning; data complexity; statistical modeling; artificial intelligent*

I. INTRODUCTION

Bigdata means datasets whose size is beyond the ability of typical database software tools to capture, store, manage, and analyze. Bigdata indicates the capability of using entire data sets instead of just segments as in years past. Bigdata era is that companies, governments and nonprofit organizations have experienced a shift in behavior. More data is a buried treasure waiting to be discovered by curious [1]. Actually, the bigdata always exists but just for those professional company or government before and it comes to be public. Bigdata is far more powerful than the analytics of the past and it makes better predictions and smarter decisions possible [2]. The use of the bigdata and analytics has become commonplace and far-reaching. Companies across the globe and in a multitude of industries see data and analytics as a mean for innovation, operational efficiency and future success [3]. The bigdata phenomenon is changing our world. The framework offered within this work reveals how practices firms are leveraging bigdata initiatives to develop and sustain competitive advantage in varying degrees. Executives should align their bigdata aspirations to get on the right track and study ahead of the curve on innovation, competition and productivity [3].

Bigdata has the attention of most every industry, with executives across the globe seeking guidance for best practices and greater understanding of the bigdata should play in strategic decision making [3]. Bigdata has significant implications for the business community, as bigdata is big business on a global scale. The Economist recently declared that digital information has overtaken oil as the world most valuable commodity [4]. The Gartner estimates that total software, social media, and IT services spending related to bigdata and analytics topped 28 billion USD worldwide in 2012 [5]. Global data flows accoutered for a 2.8 trillion USD increase in global GDP in 2014 [4].

The magnitude and power of the bigdata is more in-depth than anything which are ever experienced before and the bigdata has three key differentiating factors [2]:

- **Volume:** The amount of data that crosses the internet every second in 2010s is more than what was stored (Exabytes or Petabytes) in 1990s.
- **Velocity:** Speed of incoming data is often more important than volume, as speed determines how ahead you are of competitors.
- **Variety:** Bigdata pulls together data from a variety of channels, that are relativity new and constantly changing.

Data scientists realize that they face technical limitations, but they do not allow that to bog down their search for novel solutions. They advise executives and product managers on the implications of the data for products, processes, and decisions [6]. Companies across the globe increasingly have come to the realization that the ability to analyze and use big and complex datasets are the most important source of competitive advantage in the 21st century [7]. Yahoo was instrumental in developing Hadoop. Facebook data team created the language Hive for programming Hadoop projects. Many other data scientists, especially at data-driven companies such as Google, Amazon, Microsoft, Walmart, eBay, LinkedIn, and Twitter, have added and refined the tool kits [6]. As it is shown, many companies are investing like crazy in data scientists, data warehouse, and data analytics software.

---

Unfortunately, even global companies do not have much to show for their efforts [5]. The biggest reason that investments in the bigdata fail to pay off is that many companies are not ready to embrace the leverage of bigdata for improving the company performance, as it requires overcoming a number of barriers including developing new employee skills and upgrading IT infrastructure, instilling new management practices and new company culture across the entire company [7]. They do not know how to manage, to analyze and to understand the data [5]. More importantly, they do not know how to construct the compact data from the bigdata. Bigdata is big business and all estimates are expecting rapid growth. It is noted that analyzing the impact of business rules does not require the massive processing and the statistical modeling associated with bigdata.

Bigdata is the aggregation of large-scale, voluminous, and multi-format data streams originated from heterogeneous and autonomous data sources [8]. On the other hand, bigdata inherits the curse of dimensionality and millions of dimensions are required to be effectively reduced to be compact data [8-11]. Although the bigdata bring new opportunities to modern society, the bigdata introduces unique computational and statistical challenges [9]. To handle these challenges, it is urgent to develop the compact data by using data reduction methods including statistical methods that are robust to data complexity, noises and data dependence [9].

Compact data design is a conceptual idea to design an optimized dataset that gives best assets without handling complex bigdata. Hence, the compact data should contain the maximum knowledge patterns at fine-grained level for effective and personalized utilization of bigdata systems [12,13]. The common ways to develop the compact data [8] are as follows:

- **Network Theory** is playing a primary role in reduction of high-dimensional unstructured bigdata into low-dimensional structured data [14].
- **Compression:** The reduced-size datasets are easy to handle in terms of processing and in-network data movement inside the bigdata storage systems.
- **Data Deduplication:** Data redundancy is the key issue for data analysis because of adding nodes, expanding datasets and the data replication.
- **Data Preprocessing** is the second important phase of bigdata processing which should be preprocessed before storage at large-scale infrastructures [15].
- **Dimension Reduction** is mainly considered because the massive bigdata collection and computational complexity [16].
- **Data Mining** is required because huge efforts and computational power to explore the incrementally growing uncertain search space.
- **Machine Learning** models are inherently complex and some recent learning approaches [17,18].

Design of proper compact dataset is especially vital for developing the artificial intelligent (AI) and the machine learning (ML). The Convolutional Neural Network (CNN) is one of popular machine learning models for various computer vision applications [19]. The CNN performance improvement highly depends on the processor architecture when deploying these redundancy removal methods on existing embedded devices. A kernel pruning approach removes kernels with high sparsity to reduce the computation [20, 21]. A combination of algorithmic advancements has led to a breakthrough in the effectiveness [22]. This research provides innovative and unique ways to design compact data. Developing compact data is not only included in down-sizing data but also in a whole process to handle compact dataset. Recently, practical applications for using the compact data are introduced. Various compact data techniques have been applied into the wide range of data-driven research area including the machine learning based biometrics [23-25], the statistical civil engineering analysis [26] and developing the COVID-19 epidemic predict model [27].

The paper is organized as follows: Section II provides the practical cases which the compact data model has been applied into real-world problems. This section contains empirical studies to show how compact data could be properly designed. The actual compact data methods are briefly in Section III. Finally, the conclusion is included in Section IV.

## II. APPLICATIONS OF COMPACT DATA DESIGN

This section provides empirical studies of which compact data methods have been applied into real-world research. The performance of the machine learning engines could be improved simply by redesigning input attributes. The compact data method is a tailor-made design which depends on problem situations.

### A. Machine Learning Applications in Biometrics

Biometric authentication is replacing typical complex identification and access control systems to become a part of everyday life [24]. The time slicing technique is considered for building up the dataset of the machine learning training [23]. This approach is especially applicable for building up the machine learning training dataset The ECG data are sliced based on a window time with the R-peak anchoring (see Figure 1).

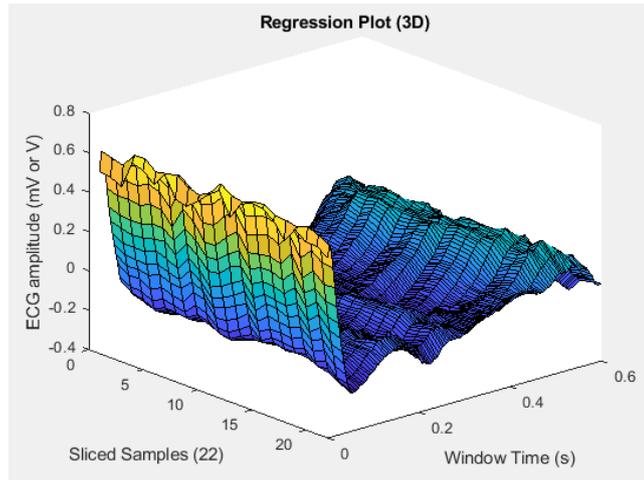

Figure 1. ECG time slicing with R-peak anchoring [24]

This method could generate enough data samples and each sliced data becomes a sample input for the machine learning training. Sliced ECG data in the time domain is chopping ECG signals starting from each R-peak moment to the sliding window period and layering these pieces based on the R-peak moment. The time sliced ECG method has been applied to build the training and testing datasets [24].

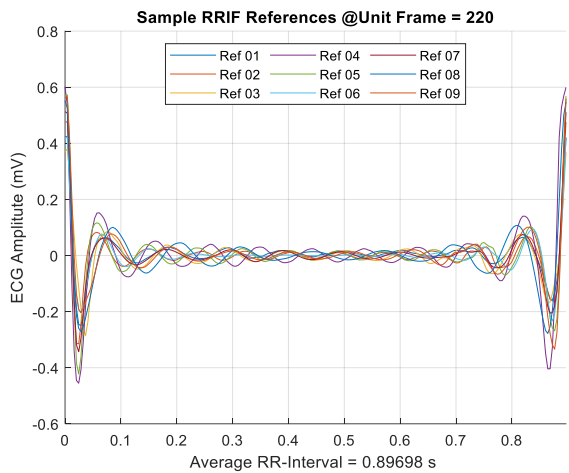

Figure 2. RR-Interval Framed ECG [25]

Alternatively, slicing the ECG data based on the interval between heartbeats (RR-Interval) instead of the slicing window time is the basis of the RR-interval framing (RRIF) method [25]. The RRIF provides is highly effective for ECG analysis and it has been applied for an ECG based authentication system by using compact size of data which feed the machine learning engine [25]. The time sliced ECG compact data could be adapted into IoT devices such as smart watches and smartphones for authentication of device owners.

*B. Composites Detection by HNSWs*

The compact data design could be also applied for defect detection in laminated composite materials using Highly Nonlinear Solitary Waves (HNSWs) based nondestructive evaluation technique [26].

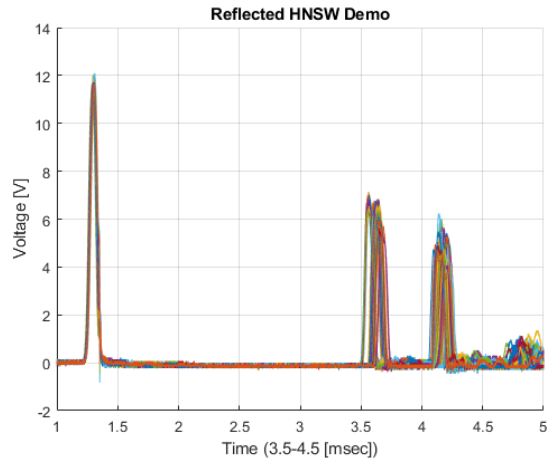

Figure 3. Data collection of HNSWs [26]

A CNN is employed to process HNSWs data collected from a granular crystal sensor [28]. To improve the efficiency of the proposed Deep-Learning (DL) algorithm, this research uses the time sliced HNSWs including only the reflected HNSWs within the range (see Figure 3). These time sliced HNSWs data are then used for training, testing and validation of the CNN system. This approach has provided a new way of building the input parameters for DL training and the detection of defects in AS4/PEEK composites without explicitly analyzing characteristic features of the reflected HNSWs (see Figure 4).

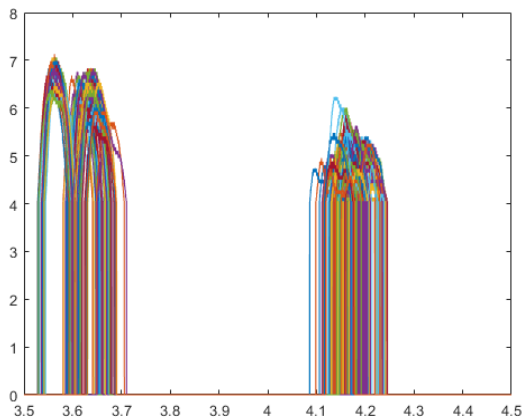

Figure 4. Fixed time sliced HNSWs data [26]

*C. Advanced Analytical Epidemic Diffusion Model*

The Advanced Analytical Epidemic Diffusion Model (AAEDM) is a dynamic diffusion prediction model which is theoretically intuitive and its tractable closed formula could be easily adapted into versatile Bigdata driven analytics including the machine learning system [27]. This dynamic model is still an analytical model but the periods of prediction are segmented for adapting the values from the dataset when the data is available. This study has proposed two important factors (i.e., kappa-factor and zeta-factor) which determine the characteristics of the epidemic diffusion (see Figure 5). Although this analytical model has been designed from a basic exponential growth model, the performance of the AAEDM is competitive even with other bigdata based simulation models.

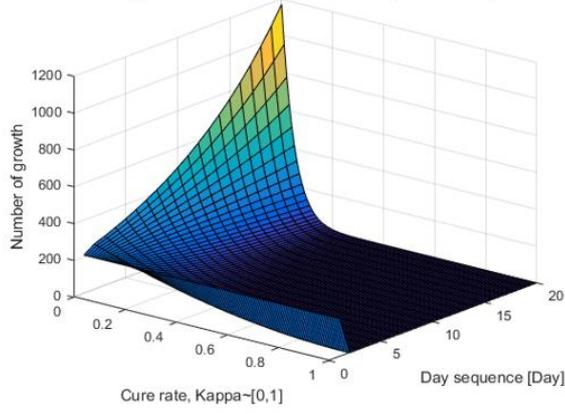

Figure 5. 3D graph for the kappa-factor in AAEDM [27]

### III. ADVANCED STATISTICAL METHODS FOR COMPACTIZING DATA

#### A. Some Performance Measures for Machine Learning Based Biometrics

Evaluating machine learning algorithms is an essential part of any ML projects and delivering good quality samples is vital for the ML algorithm evaluations [23-25]. The tailor-made process by using a tailor-made compact data design is capable to minimize workload of a performance measuring process. The *Mean Absolute Error Rate (MAER)* [23] is defined as follows:

(1)
$$MAER = \frac{1}{N}\sum_{n=1}^{N}\frac{|Y_n - \mu_n|}{\mu_n + \varepsilon}, \varepsilon \sim 0,$$

where $\mu_n$ is the reference (typically the mean) value of the data $Y_n$. It is noted that the formula (1) has been designed to avoid for dividing zero. The reference values ($\mu_n$, n = 1, ..., N) in the MAER are the mean values of the prediction after the machine learning training.

The *Accuracy Percentage within Ranges (APR)* [23] indicates the quality of the time sliced ECG data even before validating data and the larger *APR* indicate the better performance. The ARP is a portion of a ECG dataset within the ranges between 0 and the upper control limit [23].

(2)
$$APR = \frac{n(samples\ within\ the\ UCL)}{n(total\ samples)}.$$

The *Overall Performance (OP)* [25] gives an overall performance of an authentication system based on the accuracy but also the data quality. The OP is defined as follows:

(3)
$$\Pi = \left(\frac{\varphi}{N}\right) \cdot \chi$$

where φ = the number of accepted data samples, *N* = the total number of data samples and χ = the accuracy determined using the confusion matrix (0 ≤ χ ≤ 1). This performance measure could be parallelly used for analyzing machine learning systems.

#### B. Revised Mode Based Standardization

This compact design method deals the new approach about an average mean and an average variance of data samples. These two factors are classically applied in wide range of the statistics and the quality engineering. Currently, these two conventional factors are applied in the machine learning and the data sciences [29,30]. Let us assume the dataset X = $\{X_1, X_2, ..., X_n\}$ and each $X_k$ is iid (independently identically distributed) with the mean µ and the variance σ, the average mean and the standard variance are defined as follows:

(4)
$$\bar{X} = \frac{\sum_{k=1}^{n} X_k}{n}, \lim_{n \to \infty} \bar{X} = \mu,$$

$$\bar{s}^2 = \frac{\sum_{k=1}^{n}(X_k - \mu)^2}{n}, \qquad \bar{s} = \frac{\sqrt{\sum_{k=1}^{n}(X_k - \mu)^2}}{\sqrt{n}} \tag{5}$$

Most statistical data analysis uses the above formulas for finding errors, optimums and the standardized (or normalized) values. The standardization means adjusting values measured on different scales to a normal scale, often prior to averaging. In the statistics, a random variable $X_k$ is standardized by subtracting its expected value and dividing the difference by its standard deviation [31]:

$$Z = \frac{X_k - \mu}{\frac{\sigma}{\sqrt{n}}}, \mu \simeq \bar{X}, \sigma \simeq \bar{s}. \tag{6}$$

The *Mode Based Mean* and *Variance* are suggested instead of using the average. The mode refers to the most frequently occurring number found in a set of numbers [32] but it may not effective enough to use when the portion of the mode are not high enough. Therefore, the revised mode is proposed which could be formalized as follows:

$$\varphi = \frac{1_{\{\hat{p} \geq \eta\}}}{\hat{p}} \cdot E[X \cdot \mathbf{1}_{\{X=x^*\}}] + 1_{\{\hat{p} < \eta\}} \cdot E[X] \tag{7}$$

where

$$\hat{p} = \underset{p}{\mathrm{argmax}}\, p(x), p(x) = P\{X = x\}, X \in \Omega \to x \in R_+ \tag{8}$$

and $\eta \in [0,1]$ is the threshold probability that the mode is considered as a representative of random variables instead of the mean. The revised mode becomes the mean if the portion of the most frequent data is smaller than (i.e., $\eta < \hat{p}$). Similarly, the *Mode Based Variance* could be defined as follows:

$$\hat{\sigma}^2 := E[(X - \varphi)^2] = E[X^2] - 2E[X] \cdot \varphi + \varphi^2 \tag{9}$$

and

$$\hat{\sigma}^2 = E[X^2] - 2\mu \cdot \varphi + \varphi^2 \tag{10}$$

where

$$E[X^2] = \frac{\sum_{k=1}^{n}(X_k)^2}{n} \tag{11}$$

From (6), (7) and (9), the *Mode Based Standardization* is

$$W = \frac{X - \varphi}{\frac{\hat{\sigma}}{\sqrt{n}}}. \tag{12}$$

## IV. CONCLUSION

Bigdata has the attention of most every industry with executives across the globe seeking guidance but it still contains data complexity, noises and data dependence. The compact data is the revised data for optimizing the data handling process and the tailor-made compact data design has been adapted even into IoT devices as the parts of the machine learning implementation. Compact data is an optimized dataset that gives best assets without handling complex bigdata. This research provides innovative and unique ways to design compact data.